\documentclass[11pt]{article}
\usepackage{acl2014}
\usepackage{times}
\usepackage{mathtools}
\usepackage{booktabs}
\usepackage{graphicx,url}
\usepackage{natbib}
\usepackage{hyperref}
\usepackage{multirow,tabularx}
\usepackage[english]{babel}   
\usepackage[utf8]{inputenc}  
\usepackage{lmodern}
\usepackage[T1]{fontenc}
\usepackage{textcomp}
\usepackage{listings}
\usepackage{color}
\usepackage{url}
\usepackage{latexsym}

\title{Complex Networks Analysis for Software Architecture: an Hibernate Call Graph Study}

\author{
\begin{tabular}[t]{c@{\extracolsep{1em}}c@{\extracolsep{1em}}c} 
Daniel Henrique Mourão Falci  & Orlando Abreu Gomes & Fernando Silva Parreiras \\
danielfalci@gmail.com & orlando.gomes@fumec.br & fernando.parreiras@fumec.br \\
\\
\multicolumn{3}{c}{LAIS -- Laboratory for Advanced Information Systems} \\
\multicolumn{3}{c}{FUMEC University}\\
\multicolumn{3}{c}{Av. Afonso Pena 3880 - 30130-009 - Belo Horizonte - MG - Brazil}\\
\\
\end{tabular}
}

\date{}

\begin{document}
\maketitle
\begin{abstract}
  Recent advancements in complex network analysis are encouraging and may provide useful insights when applied in software engineering domain, revealing properties and structures that cannot be captured by traditional metrics. In this paper, we analyzed the topological properties of Hibernate library, a well-known Java-based software through the extraction of its static call graph. The results reveal a complex network with small-world and scale-free characteristics while displaying a strong propensity on forming communities.
\end{abstract}

\section{Introduction}

\begin{table*}[htbp]
  \centering
  \resizebox{\textwidth}{!}{%
  \begin{tabular}{@{} clcl @{}}
  \toprule
    \textbf{Caller type} & \textbf{Caller instruction} & \textbf{Callee type} & \textbf{Callee instruction}\\ \midrule
    M & org.hibernate.cfg.Configuration::setProperty & M & java.util.Properties::setProperty\\
    M & org.hibernate.cfg.Configuration::addPackage & M & org.hibernate.boot.MetaDataSources::addPackage \\
    M & org.hibernate.cfg.Configuration::registerTypeContributor & I & java.util.List::add \\
    M & org.hibernate.cfg.Configuration::registerTypeOverride & O & org.hibernate.type.CustomType::new \\
    C & org.hibernate.cfg.Configuration & C & org.hibernate.type.CustomType \\
    \bottomrule
  \end{tabular}
  }%
  \caption{CallGraphExtractor sample output.}
  \label{tab:callgraphoutput}
\end{table*}


In the Software Engineering field, one has been observing an unceasing search for quantitative measures capable of assessing internal software attributes such as maintainability, reusability, agility, among others. The argument is that these characteristics, also known as architecturally significant requirements (ARS), when combined, determine the product quality \citep{chen2013characterizing}, what in turn affects the development costs, particularly while under the software maintenance cycle. According to \citet{nelson2005survey}, this stage responds for a share between 50\% and 80\% of the total development cost.

The first measurement processes were based on metrics that are relatively simple to obtain such as the amount lines of code, cyclomatic complexity and commentaries percentage on source code \citep{hudli1994software, lee1994some}. With the strengthen of object-oriented programming paradigm and due to the high complexity of nowadays systems, a set of metrics focused on describing the association levels among software components arose. In this sense, we highlight encapsulation, coupling, and cohesion \citep{allen1999measuring, mitchell2003concepts}. These measures are particularly useful for the Software architecture, a sub-field of Software Engineering that investigates how to best to segment a system into independent and intercommunicating components, favoring the software attributes mentioned above \citep{shaw1995abstractions}.

A natural choice to study the association level among software components is through the usage of call graphs, where vertexes may represent software elements in a system, and the edges map their calls, giving shape to a complex network of relationships \citep{pezze2009teste}. Such representation may be dynamic, indicating that it was captured during system execution, or static that on the contrary, analyzes the source code structure. Call graphs have proven its utility in many software development activities such as compiler optimization and program understanding \citep{graham198gprof, bohnet2006visual}.

The complex networks theory field, in turn, boosted by the sharp increase in computing power in recent years, has evolved significantly. Its new methods and techniques created a favorable environment for the development of new approaches applicable on pattern-discovery, community and anomaly detection, trend prediction and others in several fields, from social to biological sciences \citep{watts1998collective, barabasi1999emergence, newmannetworks, blondel2008fast}.

Considering this scenario, the following research question emerge: "Which software attributes may be revealed through the application of common complex network analysis measures on call graphs?". We also intend to analyze the topological and basal properties of software in network theory field. In this work, we utilize the Hibernate library, a well-known Object/Relational Mapping (ORM) framework, widely employed in Java-based enterprise applications.

The rest of this paper is organized as follows: In section 2 we present the related work. In section 3, we expose our methods and materials. In section 4 we discuss our results, and finally, in section 5 we conclude our work.

\section{Related work}

\citet{valverde2003hierarchical} analyzed an extensive collection of networks derived from object-oriented software diagrams and call graphs. In their work, the authors identified that all analyzed systems displayed small-world and scale-free characteristics, claiming that encapsulation - a programming technique that restricts the direct access to internal software components, isolating its internal complexity from the rest of the system \citep{mitchell2003concepts} - is the leading cause of such behavior. They also suggest that all object-oriented systems may exhibit the same behavior.

\citet{bhattacharya2012graph} analyzed the evolution of eleven popular open source systems, with the purpose of developing error severity predictors based on static call graphs along with staff collaboration graphs. This study proposed a measure known as NodeRank, a derivation of PageRank, which generates a rank of software components whose bugs are prone to higher severity.

Operational systems were also analyzed under the light of network theory. Through static call graphs, the source code of Linux kernel (version 2.6.27), was investigated by \citet{wang2015topology}. The authors described the topological properties of its strongly connected network portion, known as the giant component. In their study, they have found small-world and scale-free characteristics. \citet{wang2009linux} on the other hand, analyzed the evolution of filesystem and driver modules. In their study, they captured the static call graphs for 223 versions of such modules applying common network metrics in a timeline, while proposing a metric to find major structural changes.

\citet{ying2012topology} performed a topological investigation of static call graphs in a Java program, focusing on the centrality measures. This study though, does not describe any direct association to the software engineering field.

\section{Materials and methods}

This paper relied on the static call graph representation extracted from Hibernate library in its version 5.1.3\footnote{We analyzed only the \emph{core.jar} file, distributed along with the full version of Hibernate and downloadable at \url{http://www.hibernate.org}}. All the undertaken analysis presented here utilized the software Gephi\footnote{Available at \url{https://gephi.org/users/download/}} and the complex network analysis package named NetworkX\footnote{More information at \url{https://networkx.github.io}}, for the Python language. In general, we chose Gephi to create the network's visual representations and to acquire its basal properties.  NetworkX, on the other hand, was selected to perform data manipulations and any deeper investigations.

The following subsections will expose the procedures employed to create our call graph representation and the common metrics utilized in this study.

\subsection{Call graph extraction}

To extract the static call graph structure from Hibernate, we developed a tool named CGE\footnote{The source code is publicly available at \url{https://github.com/anonymous/softwarename}}. This tool is capable of reading the bytecode of Java classes embedded into JavaArchive files (JAR files) with the goal of extracting the caller and the callee for all instructions of each class contained in the referred file. Grossly speaking, our software analyzes internal methods of a class, regardless of their visibility (public, private, static, and so on), creating a relationship table in an output file. 

Table \ref{tab:callgraphoutput} illustrates the CGE output. The caller and callee types must assume a value in the following set: \emph{M} that stands for a method, \emph{O} for an object, \emph{I} for an interface, \emph{S} for a static call and \emph{C} for a class usage relationship. Based on standard {Java 8\textsuperscript{TM}} call notation\footnote{More information at \url{https://docs.oracle.com/javase/tutorial/java/javaOO/methodreferences.html}}, the instruction fields value indicate the full qualified name of a class and, when available, its respective method separated by the <::> symbol. Therefore, observing the first table row, we say that the method \emph{setProperty} from \emph{org.hibernate.cfg.Configuration} class calls the method \emph{setProperty} that belongs to the class \emph{java.util.Properties}.

When applied to Hibernate source code, the CGE extracted a relationship table containing 143,768 calls.

\subsection{Graph modeling}

The call graph structure exemplified above may serve as the raw material to several graph representations. Its vertex set may contain only objects or methods, or objects and methods combined, using directed or undirected edges, weighted or not, in a \emph{k}-partite representation. After experimenting some of these combinations, we decided to take a directed representation whose edges are unweighted, and vertexes contain methods and also objects in a unipartite graph. 

\begin{figure*}[htbp]
    \centering
    \includegraphics[width=1\linewidth]{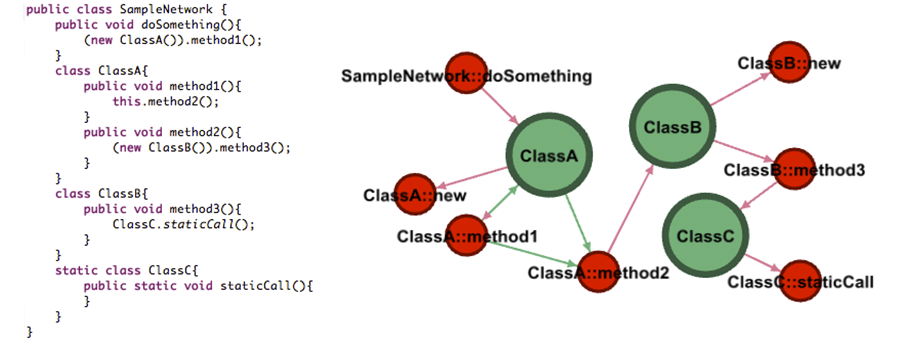}
    \caption{Resulting graph from a sample java source code}
    \label{fig:sampleGraph}
\end{figure*}

In our method, for each extracted call, we create three vertexes: The caller method, the callee class, and the callee method. If the caller and callee classes are different, indicating an object transition, we link them with two edges observing the rule: \emph{[(callerMethod, calleeClass), (calleeClass, calleeMethod)]}. If the caller and callee classes are equals, denoting an internal call, then the rule applied rule is different: \emph{[(callerMethod, calleeClass), (callerMethod, calleeMethod)]}. It is worth noting that our representation cannot be considered bipartite due to the presence of edges linking same type vertexes, in our case, methods. A sample source code and our corresponding graph representation are exemplified in Figure \ref{fig:sampleGraph}. In the sample source code, for instance, the method \emph{doSomething} of class \emph{SampleNetwork} evokes the method \emph{method1} of class \emph{ClassA}. In this case we create the three vertexes: \emph{doSomething}, \emph{ClassA}, \emph{method1}. We also create two edges linking them in a structure such as \emph{[(doSomething, ClassA), (ClassA, method1)]}. 


To create our graph representation, we implemented a software\footnote{The source code is publicly available at \url{https://github.com/anonymous/softwarename}} in Python language, which takes as input the structure generated by the CGE and, using the NetworkX package, generates the graph in Graph Exchange XML Format (\emph{GEXF}) following the rules introduced in this section. This tool was applied in the static call structure captured from Hibernate and, in order to constrain the graph representation to the library domain, we discarded the calls whose caller or callee instructions did not belong to the {"org.hibernate"} package, consequently removing any call to the native Java library. Following these procedures, we obtained a graph comprised of 27,556 vertexes and 57,919 edges.

\subsection{Network measures}

In this section we provide a general overview of the metrics utilized in this study, contextualizing each of them in software architecture field.

\subsubsection{Vertex degree}

The vertex degree is one of the most fundamental measures of complex network analysis. In a undirected graph, the degree corresponds to the number of edges linked to a given vertex\footnote{Self-loops are counted twice}. In directed graphs, though, the metric is partitioned into two. The in-degree represents the number of incoming edges in a vertex while the out-degree stands for the outgoing edges.

In software architecture, this measure may be relevant since it provides means to assess the usage level of a given software component, what may, in turn, reveal overused elements or even the coupling level among components.

\subsubsection{Betweenness}

The measure known as Betweenness is a centrality measure in complex networks theory. It reveals how many shortest paths in the network pass through each of its vertexes. In this paper, we utilized the algorithm proposed by \citet{brandes2001faster} to calculate it.

We selected this measure based on the perception that it may provide a suitable representation for object encapsulation \citep{mitchell2003concepts}. The idea is that an encapsulated software component shall provide few entry points to the rest of the system, hiding its internal connections and structures. This concept leads to a higher betweeness degree for the entry components. Therefore, this measure may indicate encapsulation level of a given software element, a useful information while in software development cycle.


\subsubsection{PageRank}

PageRank is a centrality measure in complex network analysis that is commonly treated as a measure of importance. It is calculated through the analysis of the number of edges received by a vertex and by the quality of the vertexes that bind to it, which is denoted by its own \emph{PageRank}. In its calculation procedure, takes into account the out-degree of each vertex, normalizing the index \citep{newmannetworks}. In general, the method translates the notion that a vertex is important if it is connected to other important vertexes.

When applied to software architecture, this measure may be helpful in revealing key software components, which is obviously important in several activities such as software understanding and bug severity prediction.

\subsubsection{Modularity}

Modularity is a topological measure that is concerned with the propensity of a given network to form communities. Formally, it identifies network sections whose vertexes are densely connected to each other (internal to the group) and weakly connected to vertexes belonging to another vertexes group. In this paper, we utilize the algorithm proposed by \citet{blondel2008fast}, which translates this concept into a number ranging from 0 to 1, where higher values indicate a higher propensity to form communities. Another relevant result of this algorithm is that it associates each vertex with a group what may provide useful information for further processing.

In software architecture this technique may dynamically discover groups of related software structures, considering the intensity of their relationships. This fact indicates a remarkable characteristic since traditional grouping structure is typically static, package oriented and established during the artifact creation. The dynamic community detection may be useful in detecting structural problems (code smells) and patterns.

\section{Results and Discussion}

In this section we present the basal properties of Hibernate network analyzing its small-world, scale-free and community properties. Table \ref{tab:basalProperties} presents the values for some of the main complex network measures. Among them, we highlight the average shortest path (\(19.664\)), which stays on pair with the values achieved for the Linux kernel ($\approx 21$) \citep{wang2015topology}.

\begin{table}[htbp]
  \centering
  \begin{tabular}{@{} ll @{}}
  \toprule
    \textbf{Measure} & \textbf{Value}\\ 
    \midrule
        Average clustering coefficient & \(0.194\) \\
        Average shortest path & \(19.644\) \\
        Average degree & \(2.02\) \\
        Network diameter & \(62\) \\
        Modularity & \(0.838\) \\
        Number of communities  & \(446\) \\
    \bottomrule
  \end{tabular}
  \caption{Hibernate network measures}
  \label{tab:basalProperties}
\end{table}

\begin{table*}[htbp]
    \centering
    \resizebox{\textwidth}{!}{%
    \begin{tabular}{@{} clll @{}}
    \toprule
        \textbf{Rank} & \textbf{Betweeness} & \textbf{Degree} & \textbf{PageRank} \\
        \midrule
    1 & org.hibernate.type.TypeFactory & org.hibernate.internal.CoreMessageLogger & org.hibernate.internal.util.StringHelper\\
    2 & org.hibernate.boot.cfgxml.spi.LoadedConfig & org.hibernate.internal.CoreMessageLogger.\$logger & org.hibernate.internal.CoreMessageLogger\\
    3 & org.hibernate.boot.MetadataSources & org.hibernate.engine.spi.SessionImplementor & org.hibernate.internal.CoreLogging\\
    4 & LoadedConfig::consume & org.hibernate.engine.spi.SessionFactoryImplementor & org.hibernate.engine.spi.SessionImplementor\\
    5 & org.hibernate.boot.cfgxml.spi.MappingReference & org.hibernate.internal.util.StringHelper & org.hibernate.engine.spi.SessionFactoryImplementor \\
    6 & MappingReference::apply & org.hibernate.type.Type & org.hibernate.type.Type\\
    7 & org.hibernate.boot.internal.MetadataBuilderImpl & org.hibernate.persister.entity.EntityPersister & org.hibernate.internal.CoreMessageLogger.\$logger\\
    8 & MetadataSources::getMetadataBuilder & org.hibernate.persister.entity.AbstractEntityPersister & org.hibernate.type.AbstractSingleColumnStandardBasicType\\
    9 & MetadataBuilderImpl::build & org.hibernate.dialect.Dialect & CoreLogging::messageLogger\\
    10 & org.hibernate.dialect.Dialect & org.hibernate.mapping.PersistentClass & org.hibernate.boot.spi.SessionFactoryOptions\\
    \bottomrule
    \end{tabular}
    }%
    \caption{Top ranked software components}
    \label{tab:top10}
\end{table*}

In order to verify the existence of small world characteristics \citep{watts1998collective} in our Hibernate network representation, we calculated the link creation probability \emph{p} as shown by the equation \ref{equacaop}, where \emph{l} is the number of edges and \emph{n} represents the number of vertexes. With the \emph{p} value, we proceeded to the generation of a random network based the Erdős–Rényi model \citep{erdds1959random}. For the random network, we registered an average clustering coefficient \(C = 0\) and an average shortest path \(d = 3.479\). These values were obtained from the mean of the values obtained by five random networks. The value of \emph{C} is derived from the very low value of \emph{p}, which may produce random graphs with sparse connectivity, so sparse that may result in a disconnected graph, as it is the case. Through the comparison of the values obtained from Hibernate and random networks (formally reported in the equation \ref{eq:smallworld}), one may say that the Hibernate network presents small-world characteristics. Its clustering coefficient is much higher than the one of the random network, while the average shortest path in both networks is close, although the average shortest path obtained for Hibernate network is more than five times greater than the value achieved by the random network.

\begin{equation}
    p = \frac{2l}{n\left( n-1 \right)} = \frac{2(57,919)}{27,556\left(27,555\right)} \approx 0.0001525
\label{equacaop}
\end{equation}

\begin{equation}
\label{eq:smallworld}
\begin{gathered}
    C_{Hibernate} \gg C_{Random}  \quad \textrm{ and } \quad d_{Hibernate} \approx d_{Random} \\ 0,194 \gg 0  \quad \textrm{ and } \quad 19.664 \approx 3.479
\end{gathered}
\end{equation}

\begin{figure}[htbp]
    \centering
    \includegraphics[width=1\linewidth]{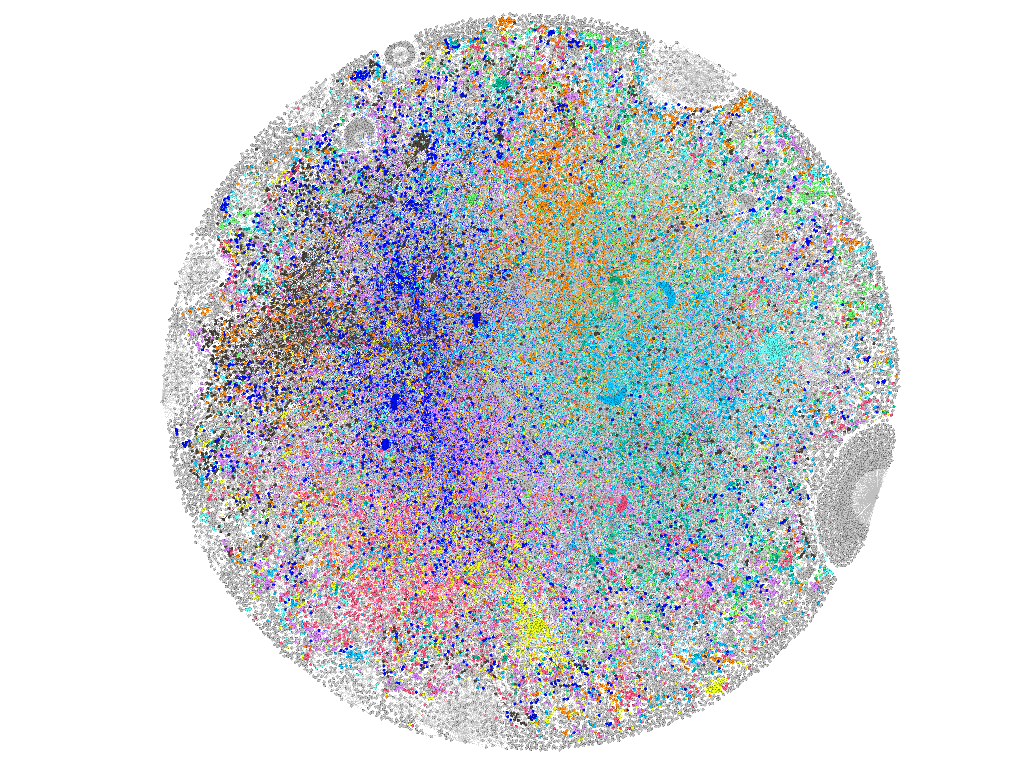}
    \caption{Network appearance - top modularity classes are colored}
    \label{fig:modularityView}
\end{figure}

\begin{figure}
    \centering
    \includegraphics[width=1\linewidth]{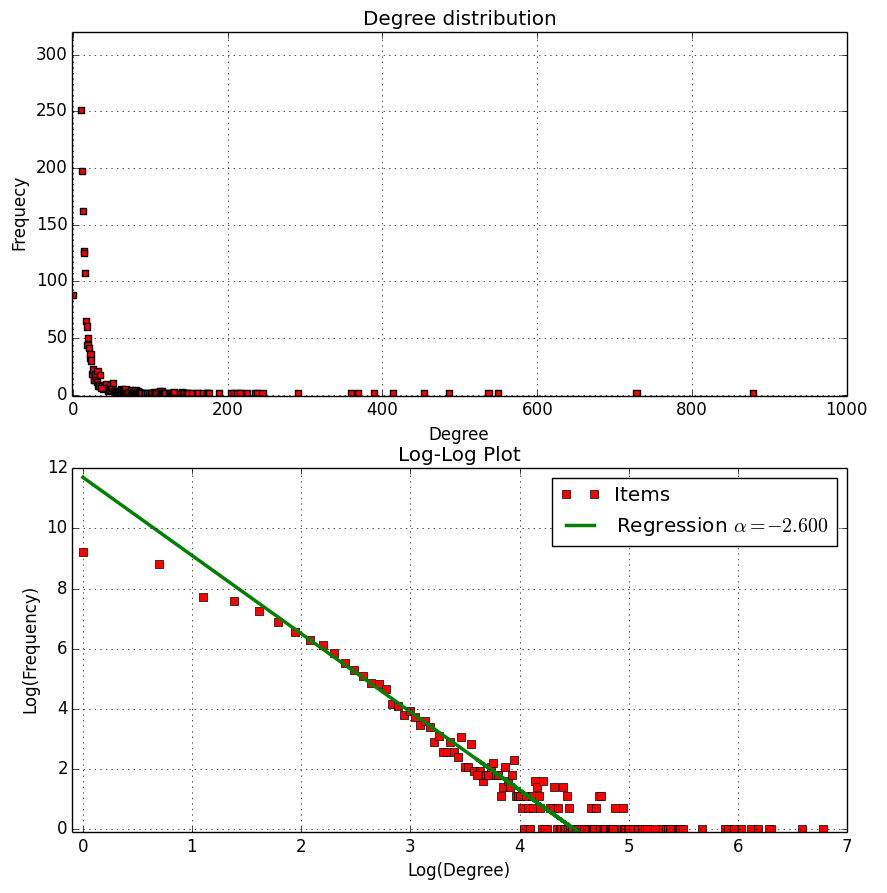}
    \caption{Scale-free analysis}
    \label{fig:scaleFree}
\end{figure}

The Hibernate network degree distribution analysis, as shown in Figure \ref{fig:scaleFree}, suggests that it may be classified as a scale-free network \citep{barabasi1999emergence}. It follows a power law distribution (\( P(k) \sim k^{-\alpha}\)) whose main characteristic is the existence of a few number of vertexes possessing a large number of connections while the vast majority of vertexes have a small number of links. The log-log plot reveal the angular coefficient \( \alpha \approx 2.6\). One of the main properties of scale-free networks is its failure resistance. This definition though, must not be used in a software context, since a failure on the smallest software component may compromise the whole system. This finding is in compliance with previous studies \citep{valverde2003hierarchical, wang2009linux, ying2012topology}. 

The modularity analysis was based on the algorithm described by \citet{blondel2008fast}, known as Louvain method. The modularity value (\(M = 0.838\)) is close to the maximum possible value (\emph{1}) which reveals a high network propensity to form communities. In fact, 446 communities were detected by this method, which results in an average value of 61.78 vertexes per class. The ten biggest classes, though, holds 47.45\% of the total number of vertexes. The Figure \ref{fig:modularityView} illustrates the aspect of the Hibernate network, where vertexes are colored according to their modularity class.

Table \ref{tab:top10} shows the top 10 vertexes obtained from the analysis of the main complex network analysis measures when applied to Hibernate's call graph. Due to the lack of comparison ground for the task of describing the relevance of such software components, particularly over the Hibernate source code structure, we limited our analysis to present the top ranked software elements for each network measure.

\section{Conclusion}


In this paper, we analyzed the properties of a widely employed Java-based software through its static call graphs. We investigated its topology that revealed a small-world and scale-free network, in compliance with the findings of \citet{valverde2003hierarchical, ying2012topology, wang2009linux}. Furthermore, we have identified that the network exhibits a strong propensity to form communities. Finally, we highlighted the top 10 software components for some of the main complex network analysis measures. 

The tools we developed for this work may benefit research in this field of study as it provides accessible means to create static call graphs for Java-based software. In future we plan to compare the traditional measures applied in software engineering  domain with the ones obtained through the application of complex network analysis measures. We also plan to investigate the application of modularity as way of finding higher level software components and its respective relevance within system.

\bibliography{acl2014}

\end{document}